\documentclass[prl,aps,twocolumn,superscriptaddress,showpacs]{revtex4}
\usepackage[T1]{fontenc}
\usepackage{ae}   
\usepackage{wrapfig}
\usepackage{float}
\usepackage{amsmath}
\usepackage{amssymb}
\usepackage{amsfonts}
\usepackage[pdftex]{graphicx}	
\usepackage{array}	
\usepackage{bm}		
\usepackage{dsfont}	
\usepackage{ulem}
\usepackage{color}
\usepackage{hyperref}
\usepackage{units}
\usepackage{gensymb}

\renewcommand{\vec}[1]{\textbf{\textit{#1}}}
\newcommand{\mx}[1]{\ensuremath{\left( \begin{matrix} #1 \end{matrix} \right) }}

\newcommand{\zref}[1]{(\ref{#1})}
\newcommand{\erw}[1]{\langle \, {#1} \, \rangle}
\newcommand{\tr}{\text{tr}}
\newcommand{\intinfty}{\int_{-\infty}^\infty}
\newcommand{\abs}[1]{\left|#1 \right|}
\renewcommand{\Im}{\text{Im}\,}

\newcommand{\e}{{\text e}}
\newcommand{\mehrzeile}[1]{\begin{cases}  #1 \end{cases} }

\renewcommand{\emph}{\textit}
\newcommand{\newcite}[1]{[\onlinecite{#1}]}

\begin{document}

\title{Tracing the electronic pairing glue in unconventional superconductors via Inelastic Scanning Tunneling Spectroscopy}
\author{Patrik Hlobil}
\affiliation{Institut f\"ur Festk\"orperphysik, Karlsruher Institut f\"ur Technologie, 76344 Karlsruhe, Germany}
\affiliation{Institut f\"ur Theorie der Kondensierten Materie, Karlsruher Institut f\"ur Technologie, 76131 Karlsruhe, Germany}
\author{Jasmin Jandke}
\affiliation{Physikalisches Institut, Karlsruher Institut f\"ur Technologie, 76131
	Karlsruhe, Germany}
\author{Wulf Wulfhekel}
\affiliation{Physikalisches Institut, Karlsruher Institut f\"ur Technologie, 76131
	Karlsruhe, Germany}
\author{J\"org Schmalian}
\affiliation{Institut f\"ur Festk\"orperphysik, Karlsruher Institut f\"ur Technologie, 76344 Karlsruhe, Germany}
\affiliation{Institut f\"ur Theorie der Kondensierten Materie, Karlsruher Institut f\"ur Technologie, 76131 Karlsruhe, Germany}
\date{\today }

\begin{abstract}
 Scanning tunneling microscopy (STM) has been shown to be a powerful
experimental probe to detect electronic excitations and further allows
to deduce fingerprints of bosonic collective modes in superconductors. Here, we demonstrate
that the inclusion of inelastic tunnel events is crucial for the interpretation
of tunneling spectra of unconventional superconductors and allows
to directly probe electronic and bosonic excitations via STM. 
We apply the formalism to the iron based superconductor LiFeAs.
With the inclusion of inelastic contributions, we find strong  evidence for 
a non-conventional pairing mechanism, likely via magnetic excitations.
\end{abstract}
\maketitle

Electron tunneling spectroscopy has turned out to be an outstanding
tool for the investigation of superconductors. A classical example
is the determination of the electron-phonon pairing interaction in
conventional superconductors~\newcite{McMillan65,McMillan68}. More recently,
quasi-particles interference spectroscopy managed to exploit the local resolution
of scanning tunneling microscopy (STM) to obtain momentum space information~\newcite{Hoffman2002,Roushan2009,Kreisel2015}.
Both examples are based on elastic tunneling theory~\newcite{Bardeen61,Cohen66}
where one interprets the low-temperature conductance to be proportional
to the electronic density of states (DOS), including renormalizations
of the DOS that occur  for example within the strong coupling Eliashberg formalism~\newcite{Eliashberg61}.
An energy dependent coupling to phonons or electronic collective modes
and the details of these bosonic spectral features lead to a renormalization
of the electronic DOS in form of peak-dip features above the superconducting
coherence peaks~\newcite{McMillan65}. Such pronounced peak-dip features have also
been observed in cuprate and iron-based superconductors~\newcite{Zasadzinski2003,Niestemski2007,Shan2012,Song2012,Chi2012,Valles91,Cucolo96,Misra2002,Nishiyama2002,Maggio1995,Renner1998,Seidel97,Matsuura98,Jandke2016b,Fasano10,Wang12}.
A frequent interpretation is, based on elastic tunneling theory, in
terms of a coupling of electrons to a sharp spin resonance mode with frequency
$\omega_{\text{res}}$ and with momentum at the antiferromagnetic
ordering vector of the material~\newcite{Eschrig2000,Abanov2000,Manske2001}.
\begin{figure}
\centering
\includegraphics[width=0.45\textwidth]{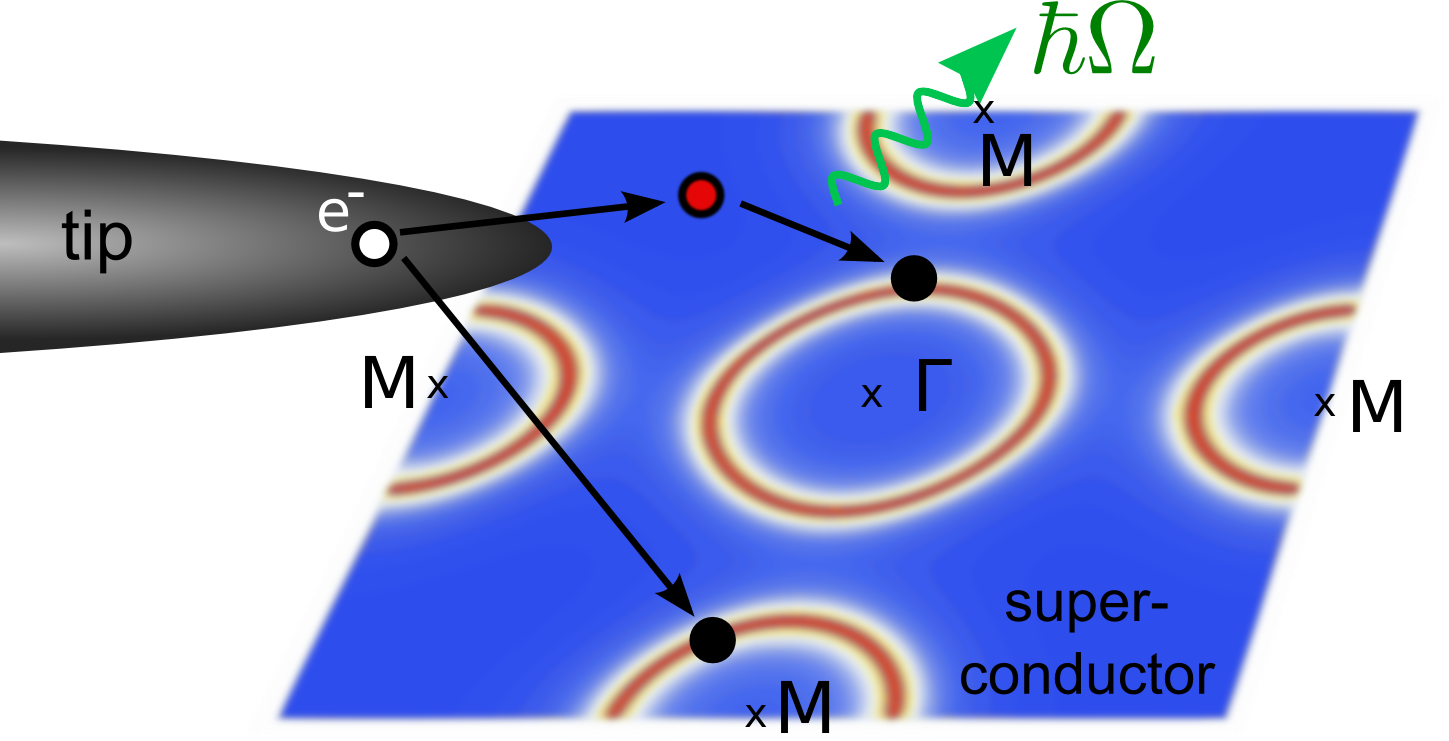}
\caption{Sketch of the elastic and inelastic tunneling processes from the tip (gray) to the band structure of the superconductor. An inelastic tunneling process involves an intermediate off-shell state far away from the Fermi surface (marked in red), from which the electron is  scattered inelastically via the emission/absorption of a boson with frequency $\Omega$ to a state near the Fermi surface (black).}
\label{Tunnelingprocesses}
\end{figure}

In an interacting system the injection of a real electron may cause
both, the creation of a fermionic quasi-particle and the excitation
of (bosonic) collective modes as depicted in Fig.~\ref{Tunnelingprocesses}. The strength of the interaction is usually crucial
for the relative weight of the low-energy quasiparticle and the cloud
of excitations associated with it. The excitation of the quasiparticle
corresponds to the above discussed elastic tunneling, while the creation
of real collective modes during the tunneling process corresponds
to \emph{inelastic tunneling}. 

\begin{figure*}
\centering
\includegraphics[width=0.3\textwidth]{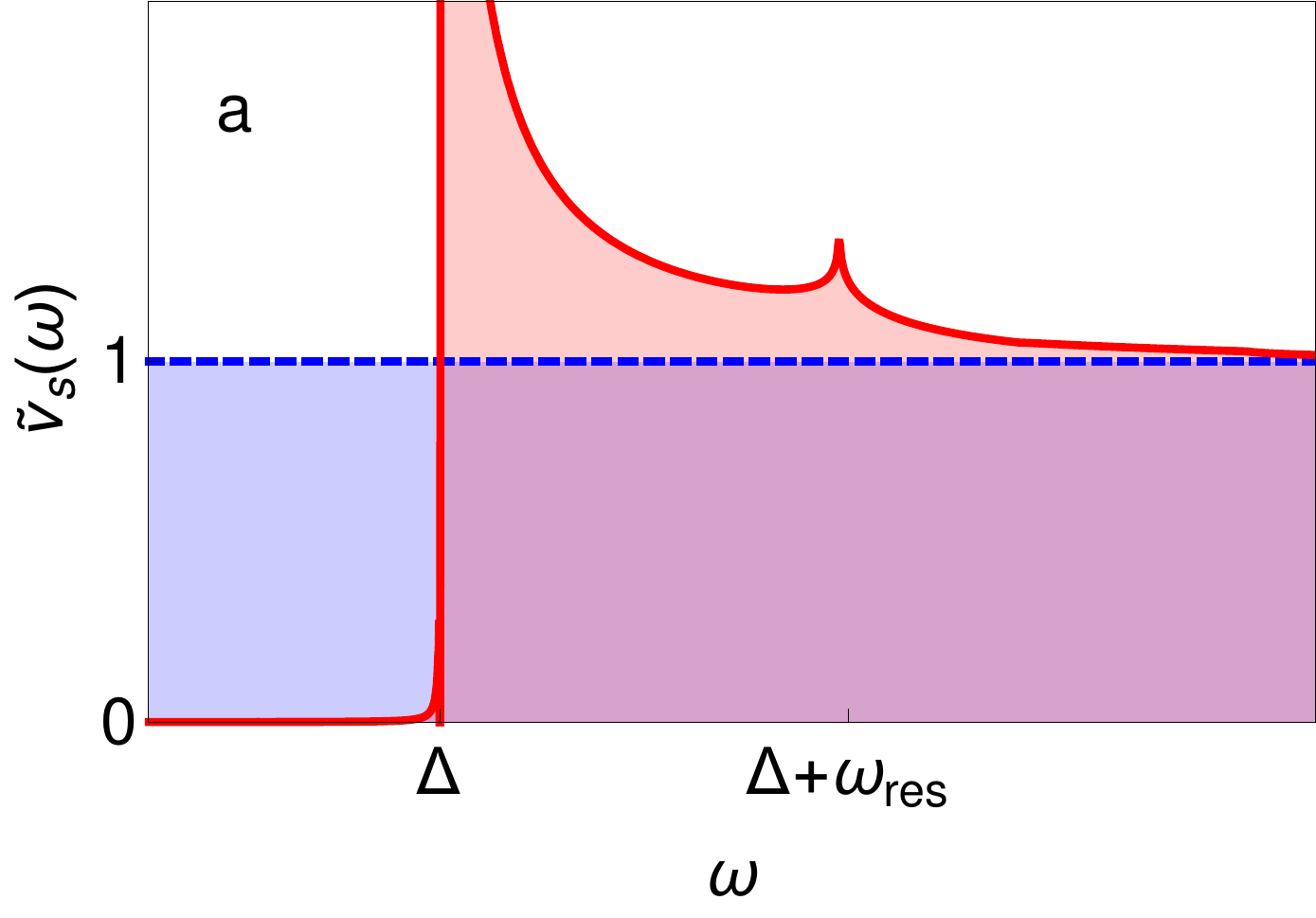} 
\includegraphics[width=0.3\textwidth]{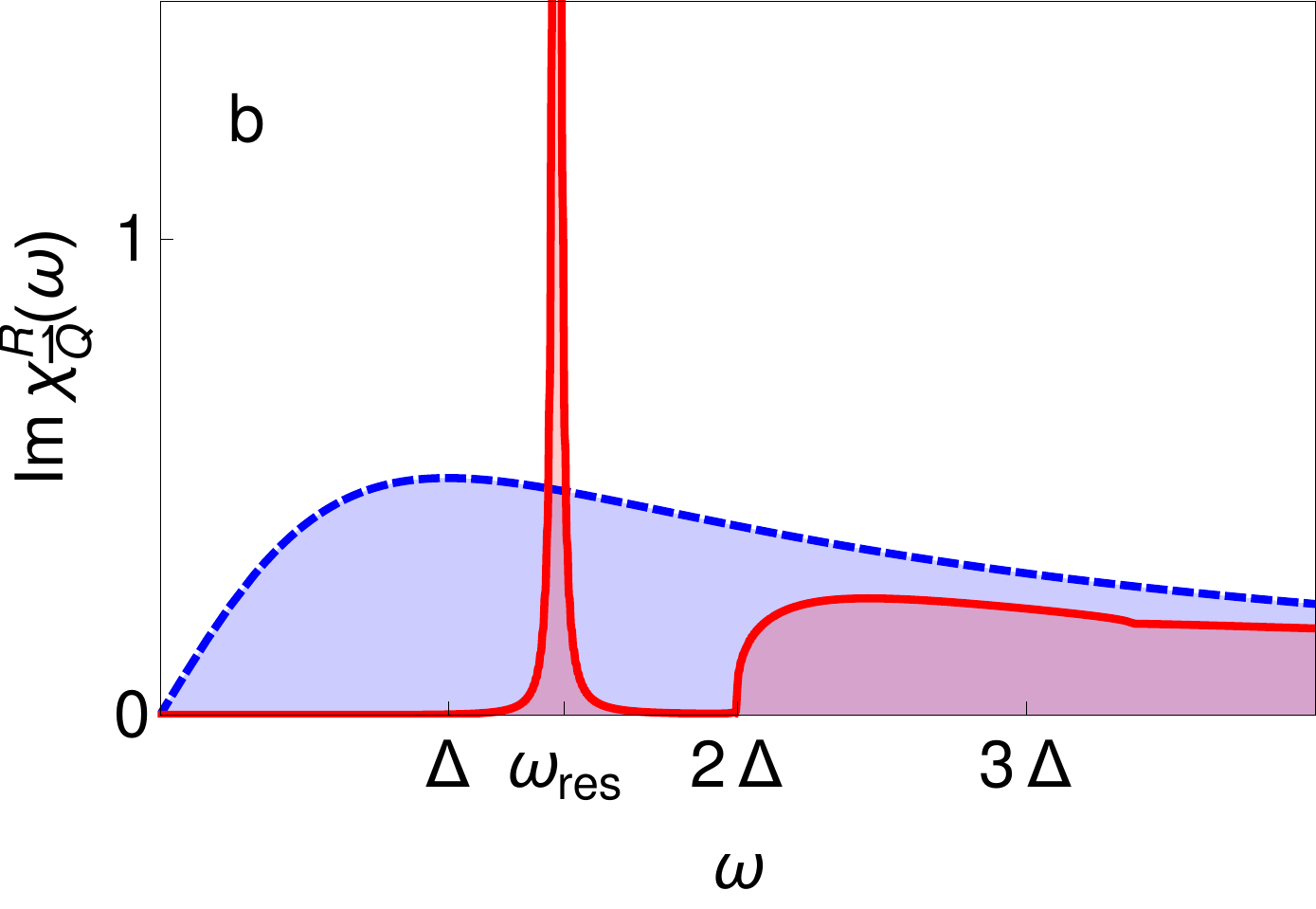} 
\includegraphics[width=0.3\textwidth]{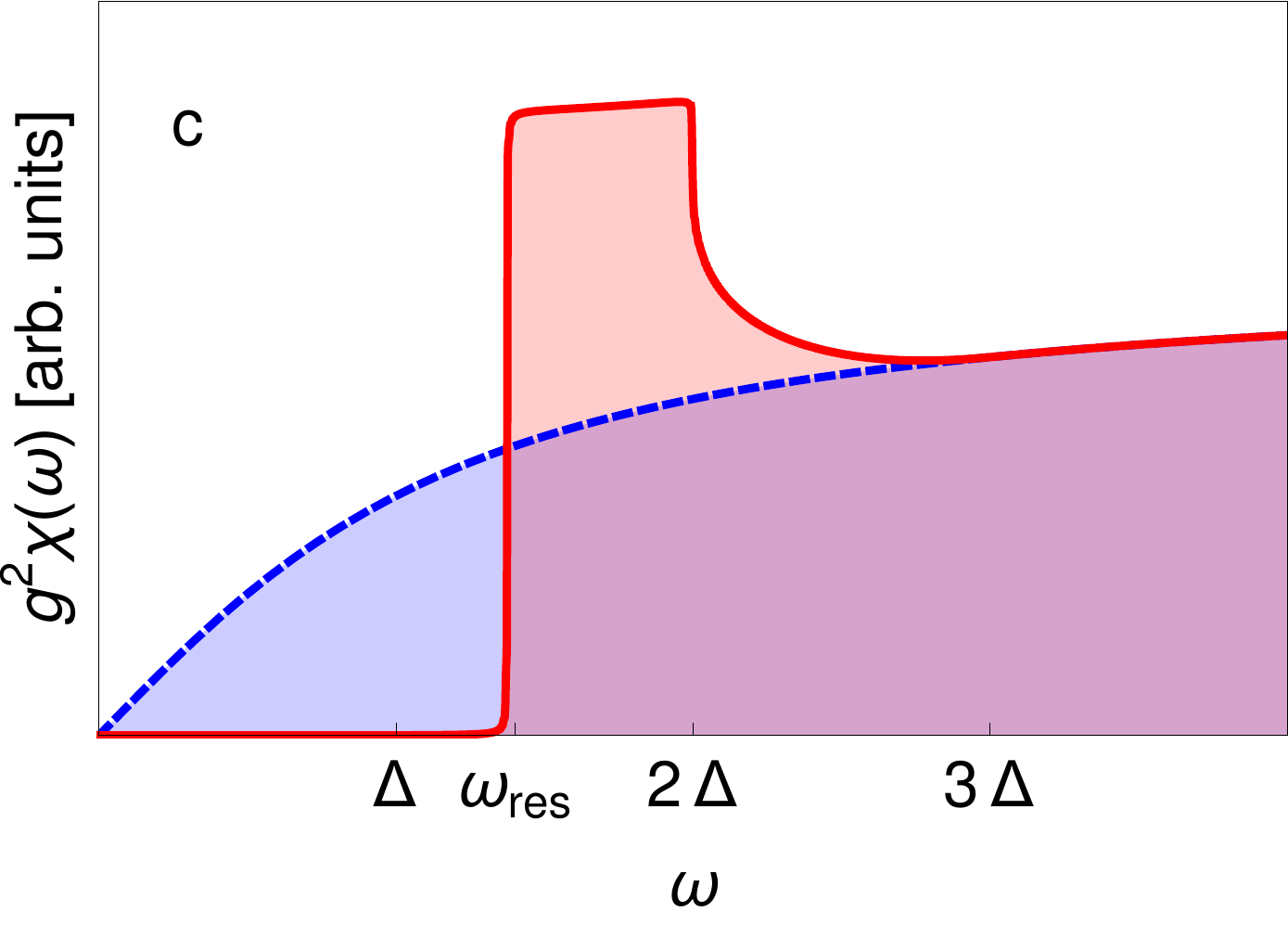} 
\caption{\textbf{Calculated spectra for the spin electron model in the  normal (blue) and superconducting state (red):} \textbf{a}, Electronic density of states. \textbf{b}, Spin spectrum $\Im \chi_{\vec Q}^R(\omega)$ at the antiferromagnetic ordering vector $\vec Q$ with the resonance mode occurring at $\omega_{\text{res}}$ below $T_c$. \textbf{c}, spin spectrum $g^2 \chi$ integrated over the 2-dimensional Brillouin zone.}
\label{Ititle0}
\end{figure*}

In this paper we demonstrate that such inelastic tunneling events
can lead to important and observable modifications of the STM spectrum
in unconventional superconductors. In addition to
fermionic exitations that are visible via elastic tunneling, inelastic
tunneling spectroscopy can be used to identify the bosonic excitations of the system. We show that the fine-structures seen in LiFeAs
 are predominantly due to such inelastic tunneling processes and thus evidence of an electronic pairing source.  Here, we analyze electronic
pairing where the  excitations causing superconductivity are directly related to the collective
bosonic modes of the electrons themselves (e.g. electron-spin fluctuations). In the superconducting state, electrons
open a gap $\Delta$ in their spectrum. This impacts all collective
excitations of the electrons. In other words, collective spin and charge degrees of freedom inherit
a gap in the  bosonic spectrum below $T_c$. This behavior is shown in the Fig.~\ref{Ititle0}, where
numerical results for the calculated electronic and spin spectral function
above and below $T_{c}$ are shown~\newcite{Bennemannbook,Abanov2001,AbanovEuro}. The spin spectrum inherits a gap $\omega_{res}$ related to a resonance mode at this energy. If the bosonic glue is made up of such a
gapped spectrum, it will strongly affect the inelastic tunneling
spectrum (much stronger than the elastic one).

An inelastic tunneling event is depicted in Fig.~\ref{Tunnelingprocesses}.
A tip-electron tunnels elastically into a high-energy off-shell state
far away from the Fermi surface from where the
electron scatters inelastically via the emission/absorption of a boson
to a state near the Fermi-surface. Inelastic tunneling has been observed
for conventional superconductors in the normal state~\newcite{Rowell69,Schackert15} where it was shown, that tunneling electrons excite bulk phonons since the measured second derivative of the tunneling turned out to be proportional to the Eliashberg function ($d^2I/dU^2 \propto \alpha ^2 F(\omega)$) which is given by the electron-phonon coupling constant $\alpha $ times the phonon spectrum $F(\omega )$. This inelastic contribution has recently been shown to be of importance even in the superconducting state of Pb-films (almost to the same extent)~\newcite{Jandke16}. Furthermore, in the normal state
of the cuprate superconductors it is well established that inelastic
tunneling channels are present and in general not negligible~\newcite{Kirtley90,Kirtley93,Littlewood92,Xiao94,Seidel97,Bennemannbook,Niestemski2007}.
They give rise to the frequently observed V-shape of the normal state
spectrum closely tied to an overdamped particle-hole spectrum as depicted
by the blue curve in Figure \ref{Ititle0}. Such V-shaped background
conductances have also been seen in the iron pnictide superconductors~\newcite{Chi2012,Fasano10,Wang12}.
In the superconducting state inelastic tunneling was discussed in
the context of fine structures of the tunneling spectrum that displayed
an isotope effect, suggesting the tunneling via apical oxygen states~\newcite{Lee06}.
We will show that inelastic tunneling below $T_{c}$  can be utilized to narrow down
the pairing mechanism in unconventional superconductors,
where one expects a dramatic reorganization of the pairing glue spectrum
in the superconducting state in contrast to the electron-phonon coupling case.

If one expands with regards to the usual tunneling matrix element
$t_{\mathbf{k},\mathbf{p}}$ between tip and superconductor\footnote{Note, that as usual STM probes the uppermost layers
of a bulk system.}, the tunneling current $I=I^{{\rm e}}+I^{\text{i}}$
consists of an elastic and inelastic contribution $I^{{\rm e}}$ and
$I^{\text{i}}$, respectively. Both are of same order in tunneling ${\cal O}\left(t^{2}\right)$, yet the inelastic contribution may be suppressed in case of momentum conservation at the tunneling junction.
Following our previous analysis for conventional superconductors~\newcite{Jandke16} we find for $t_{\vec{k},\vec{p}}\approx t$,
appropriate for STM geometries~\newcite{Giamarchi2011}, and for a constant
tip DOS the two contributions to the differential conductance $\sigma(U)=dI/dU$~\newcite{Kirtley90,Kirtley93}
\begin{widetext}{\footnotesize{}
\begin{align}
\sigma^{{\rm e}}(U) & =-\sigma_{0}\int_{-\infty}^{\infty}d\omega n_{F}'(\omega+eU)\tilde{\nu}_{S}(\omega)\ ,\label{Iel}\\
\sigma^{\text{i}}(U) & =-\frac{\sigma_{0}}{D^{2}\nu_{S}^{0}}\int_{-\infty}^{\infty}d\omega_{1}d\omega_{2}g^{2}\chi''(\omega_{1})\tilde{\nu}_{S}(\omega_{2})\biggl[n_{F}'(\omega_{2}-\omega_{1}+eU)n_{B}(\omega_{1})\bigl[1-n_{F}(\omega_{2})\bigr]+n_{F}(\omega_{2})\bigl[1+n_{B}(\omega_{1})\bigr]n_{F}'(\omega_{2}-\omega_{1}+eU)\nonumber \\
 & +n_{F}'(\omega_{2}+\omega_{1}+eU)\bigl[1+n_{B}(\omega_{1})\bigr]\bigl[1-n_{F}(\omega_{2})\bigr]+n_{F}(\omega_{2})n_{B}(\omega_{1})n_{F}'(\omega_{2}+\omega_{1}+eU)\biggr],\label{Iinel}
\end{align}
}\end{widetext}
where $U$ is the applied voltage, $\sigma_{0}=4\pi e^{2}\left|t\right|^{2}\nu_{T}^{0}\nu_{S}^{0}$
, $g$ the coupling strength between the electrons and the collective mode and $\nu_{S/T}^{0}$ are the normal state DOS of the superconductor
and the tip at the Fermi energy, respectively.  $D$ is some characteristic upper
cut off for the bosonic excitation spectrum characterized by the imaginary part of the the momentum
averaged propagator $\chi''\left(\omega\right).$ For a detailed derivation
of these expressions see the supplementary information~\newcite{Suppl}, where we demonstrate
that the distinction between elastic and inelastic tunneling is 
due to the fact that the electronic spectrum is subdivided into
a low-energy renormalized quasiparticle regime and high-energy off-shell
states. Usually, many-body interactions are analysed for the renormalized
quasiparticle excitations. However, tunneling processes into off-shell
states far away from the Fermi surface may subsequently relax into
states near the Fermi energy via the emission of a bosonic excitation.
This is a process with a large phase space as long as the typical
bosonic momentum is large. Examples are zone-boundary phonons or antiferromagnetic
fluctations. 

We also point out that the relative phase
space for elastic and inelastic processes depends sensitively on the
detailed tunneling geometry, i.e. whether one considers planar or
point-contact junctions or an STM geometry. STM settings with poor
momentum conservation~\newcite{Giamarchi2011} give large inelastic contributions. 

In the following, we investigate, for a specific model, how inelastic
tunneling affects the tunneling spectra in unconventional superconductors.
We consider the case of a spin-fermion coupling proposed as an effective
model for various unconventional superconductors~\newcite{Bennemannbook}.
The relevant collective bosonic degrees of freedom can be written
in terms of a 3-component spin vector $\mathbf{S}_{\mathbf{q}}$ with
a Yukawa-like electron-boson coupling
\begin{equation}
H_{{\rm int}}=g\int dxc_{\alpha}^{\dagger}\boldsymbol{\sigma}_{\alpha\beta}c_{\beta}\cdot\mathbf{S}
\end{equation}
with the Pauli-matrices $\sigma^{i}$.  We also define
the normalized electronic DOS $\tilde{\nu}_{S}(\omega)=\nu_{S}(\omega)/\nu_{S}^{0}$, the coupling constant $g$ and the dimensionless, integrated spin spectrum 
$\chi''(\omega)=-3\nu_{S}^{0}\int d^{d}q{\rm Im}\chi_{\mathbf{q}}(\omega)/\pi$.
We solve this model self-consistently using the formalism of Ref.~\newcite{Bennemannbook,Abanov2001,AbanovEuro} which determines the superconducting gap-function and the renormalized electron and spin-fluctuation propagators. This Eliashberg treatment is well established and the coupled set of equations is given in Ref.~\cite{AbanovEuro}. The solutions are displayed in Fig.~\ref{Ititle0}.
Recent quantum Monte Carlo calculations  confirmed that this approach is quantitatively correct as long as the dimensionless coupling constant is not much larger than unity~\newcite{Rafael}.

\begin{figure}
\centering
\includegraphics[width=0.45\textwidth]{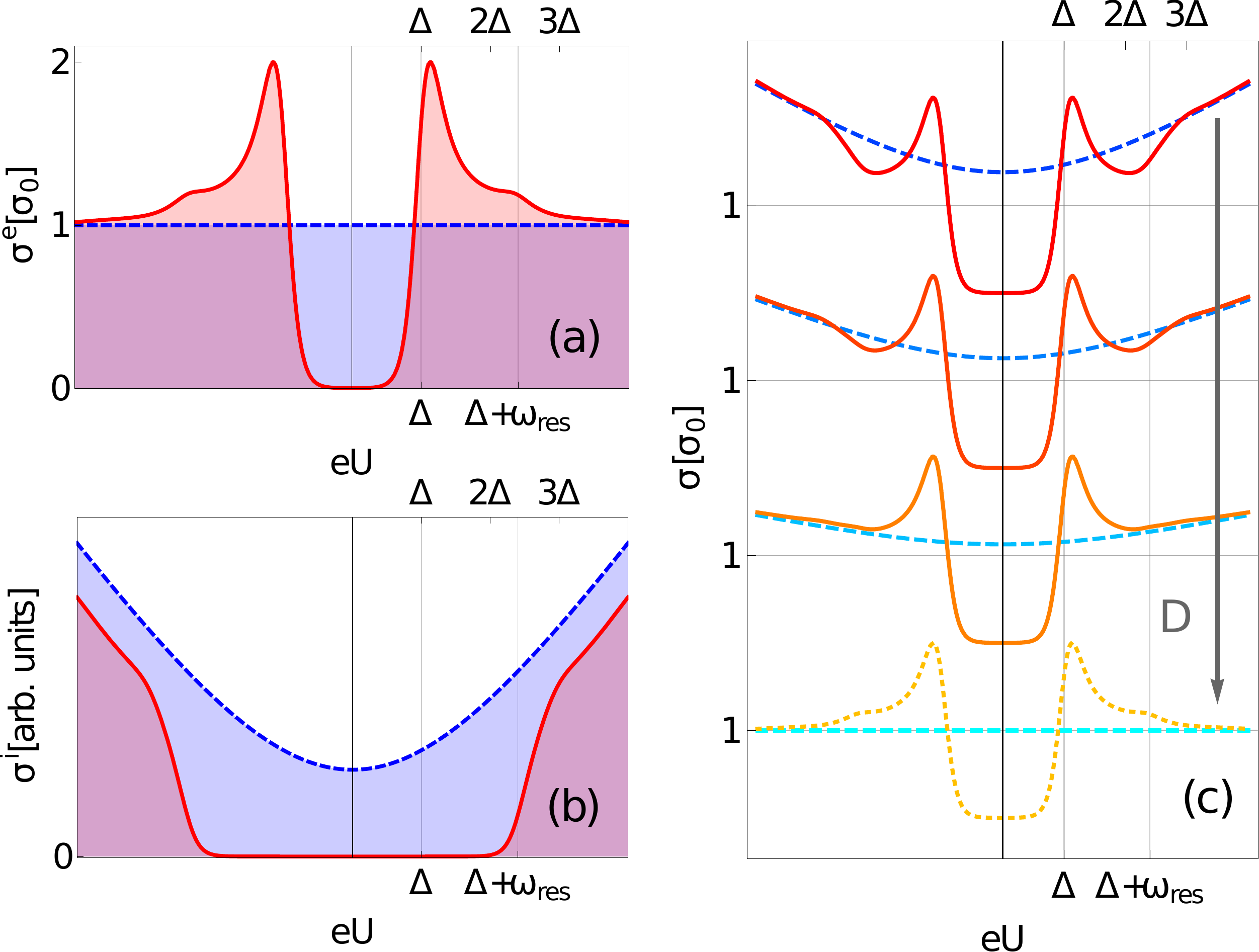}

\caption{Calculated  elastic (a) and inelastic (b) contributions to the conductance both in the superconducting (red) and normal state (blue). In (c) the total  tunneling spectra for different inelastic tunneling amplitudes  are shown. We use $1/(D^2\nu_0^S)=\unit[(0,0.2,0.5,0.8)]{1/\Delta}$ which are reasonalbel values for systems with electronic pairing. The tunneling parameters are set such that the current $I$ at $eU=10 \Delta$ is the same for the normal and superconducting state. The combination of elastic and inelastic contributions leads to the apperance of a dip feature, reflecting primarily the reorganization of the bosonic spectral weight below $T_c$ of the inelastic tunneling contribution.}
\label{Figcon}
\end{figure}

We first analyze the normal state behavior. At sufficiently low $T$ and for
a structure-less density of states,  Eq.(\ref{Iinel}) simplifies to 
\begin{equation}
\sigma^{{\rm i}}\left(U\right)\propto g^{2}\int_{0}^{eU}d\omega\chi'' \left(\omega\right).
\end{equation}
Above $T_c$ the spin susceptibility
shows an overdamped behaviour  $\chi_{\mathbf{q}}(\omega){}^{-1}\sim\xi^{-2}+(\mathbf{q}-\mathbf{Q})^{2}-\Pi_{\mathbf{Q}}(\omega)$
with $\Pi_{\mathbf{Q}}(\omega)=i\gamma\omega$, where $\gamma\sim g^{2}/v_{S}^{0}$. Here, $\mathbf{Q}$ is
the antiferromagnetic ordering vector and $\omega_{\text{sf}}=\gamma^{-1}\xi^{-2}$
the characteristic energy scale of the boson. For $d=2$ it follows
$\chi'' \left(\omega\right)=\frac{3}{2\pi}\nu_{S}^{0}\arctan\left(\frac{\omega}{\omega_{{\rm sf}}}\right)$,
which leads to $\sigma^{{\rm i}}\left(U\right)\propto g^{2}U^{2}/\omega_{{\rm sf}}$
for small voltages ($eU\ll\omega_{{\rm sf}}$) and a linear dependence
$\sigma^{{\rm i}}\left(U\right)\propto g^{2}\pi\left|U\right|$ for
$eU\gg\omega_{{\rm sf}}$, yielding a natural explanation for the
$V$-shaped (at low  voltages  rather U-shaped)  spectrum~\newcite{Kirtley90,Kirtley93}. For $T>0$, inelastic tunneling is also present at finite voltage and therefore increasing the purely elastic conductance to be larger than $\sigma_0$ at zero bias in Fig.~\ref{Figcon} (c). Note, the same can be achieved within the bosonic spectrum that underlies the marginal
Fermi liquid approach, where the role of $\omega_{{\rm sf}}$ is played
by temperature. As inelastic tunneling only probes the momentum-averaged
bosonic spectrum it cannot discriminate between these two scenarios.
Within the antiferromagnetic fluctuation theory it is however important
that the effective dimensionality of the spin-excitation spectrum
is $d=2$. For arbitrary dimension follows in the regime $eU\gg\omega_{{\rm sf}}$
that $\sigma^{{\rm i}}\left(U\right)\propto\left|U\right|^{d/2}$,
a behavior that occurs down to smallest voltages at an antiferromagnetic
quantum critical point, where $\omega_{{\rm sf}}\rightarrow0$, and may  serve to identify the effective dimension of the spin-fluctuation spectrum in a given system.
In Fig. \ref{Figcon} (a) and \ref{Figcon}(b) we show in blue the elastic and inelastic  conductance obtained from the solution of the spin-fermion model above $T_c$. While the elastic contribution is constant for the normal state, the inelastic conductance of Fig. \ref{Figcon}b)  shows the expected  V-shape structure. As discussed  earlier~\newcite{Littlewood92,Xiao94}, inelastic processes open
up additional tunneling channels for both positive and negative bias
$U$. Most important for our considerations is that the observation of an V-shaped inelastic contribution in the normal state implies that it cannot be ignored in the superconducting state and allows for an estimate of its relative contribution.

We now turn to the superconducting state. We solve
the Eliashberg equations for spin-fluctuation induced pairing numerically~\newcite{Abanov99}, considering
a nodeless pairing state (see Fig.~\ref{Ititle0}). This is appropriate for several
iron based superconductors with $s^\pm$-pairing. For systems with nodes it mostly implies
that we should confine ourselves to frequencies above the superconducting
gap, which is the regime we are interested in anyway. As usual, the
sign of the gap changes between states that are connected by the magnetic
wave vector and the resonance mode at $\omega_{res}$ naturally occurs within our formalism. 
 We have chosen our input parameters
$\omega_{{\rm sf}}$ and $g$ of the theory in such a way that
the observed gap and spin spectrum agrees well with the experimental
observations~\newcite{Bourges95,Inosov10,Christensen04,Yu2009} $\omega_{{\rm sf}}\simeq\Delta$
and $\omega_{\text{res}}\approx 1.4\Delta$, where $\omega_{{\rm res}}$
is the resonance mode seen by inelastic neutron scattering~\newcite{RossatMignod1991,Bourges1997,Yu2009,Inosov10}
that has left traces in other experimental techniques as well~\newcite{Dahm2009,Yang2009,DalConte}.
We stress that the key conclusions of our analysis are not affected by changing
the above parameters within reasonable ranges.
 
In the superconducting state, the following features arise
in the tunneling spectrum: The \textit{elastic} tunneling contribution
seen in Fig.~\ref{Figcon}(a) (red curve) is proportional to the
thermally smeared electronic DOS with the gap $\Delta$, the coherence
peak at $\Delta$ followed by the usual peak-dip strong-coupling features
seen at $\Delta+\omega_{\text{res}}$ that quickly approaches the
assumed constant DOS of the normal state for higher biases.
The \textit{inelastic} tunneling conductance seen in Fig.~\ref{Figcon}b)
(red curve) is gapped by $\Delta+\omega_{\text{res}}$ as both the
electronic DOS and the bosonic spectrum obtain a gap below $T_{c}$.
For voltages $eU>\Delta+\omega_{\text{res}}$ the inelastic differential
conductance shows a sharp increase. This behavior can be traced back to the fact that spin spectral
weight is shifted from lower to higher energies, mostly close above
the resonance mode at the frequency $\omega_{\text{res}}$. For our calculations we have chosen our temperature $T=0.1 \Delta$ in the superconducting state and $T=0.5\Delta$ in the normal state, where $\Delta$ is the gap at zero temperature.

Naturally, only the sum $\sigma=\sigma^{\text{e}}+\sigma^{\text{i}}$
is observable in tunneling experiments. 
Fig.~\ref{Figcon}(c) depicts the resulting total conductance including
the normalization for different energy cutoffs $D$. We set the tunneling parameters such that the current $I$ at
$U=10\Delta$ is the same for the normal and superconducting state. For weak inelastic
contributions ($g/D$ small) mainly the quasiparticle peak at $\Delta$
is visible on top of a small inelastic increase at high energies. Note, that the conductance in the superconducting state is always higher than in the normal state outside the gap.
The Eliashberg-features of the resonance mode are already obscured
by the inelastic contributions (second curve from below). If we increase the
inelastic tunneling amplitude we see that the conductance in the superconducting
state is lowered below the normal state at $eU\approx2\Delta$
due to the loss of spin spectral weight. Note that in the normal state, both elastic and inelastic contributions are present while in the superconducting state, there are only elastic contributions for energies  $eU < \Delta +\omega _{res}$ (see Fig. \ref{Figcon} (a) and (b)).  The obtained
spectra look similar to many measured differential conductances~\newcite{Chi2012,Valles91,Cucolo96,Misra2002,Nishiyama2002,Maggio1995,Renner1998,Seidel97,Matsuura98,Niestemski2007,Jandke2016b},
especially the fact that the superconducting differential conductance
dips lie below the normal state differential conductance above the quasiparticle
peak, followed by a V-shaped background conductance at higher energies.

\begin{figure}
\centering
\includegraphics[width=0.46\textwidth]{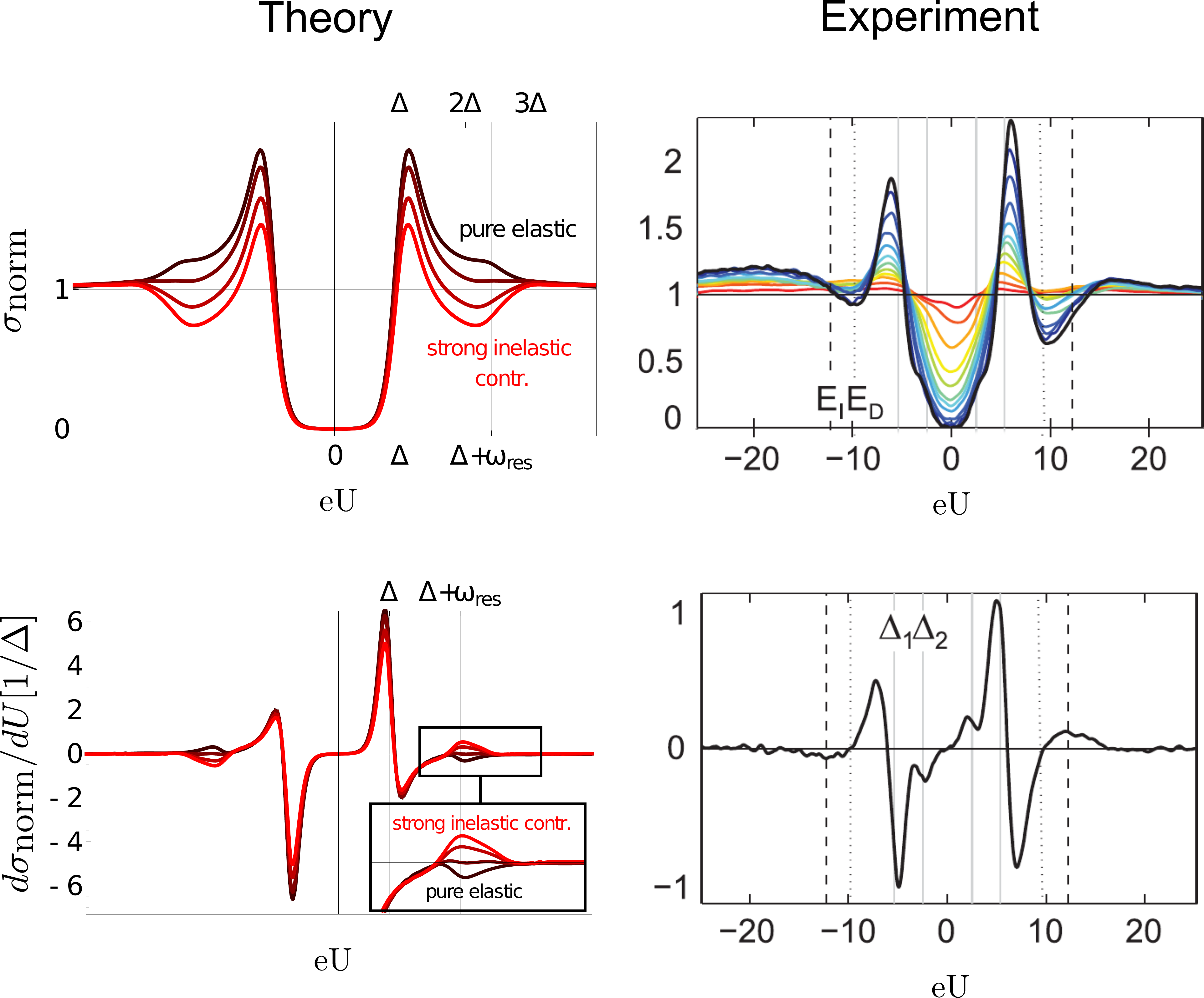}

\caption{Normalized conductance $\sigma_{\text{norm}}(U) = \nicefrac{\sigma_{\text{sc}}(U)}{\sigma_{\text{nc}}(\sqrt{U^2-(\Delta/e)^2})} $ (upper panel) and derivative of the normalized conductance (lower panel). We use the same values for $1/(D^2\nu_0^S)$ as in Fig. \ref{Figcon} (c). Left we show our theoretical results, compared to the STM data of LiFeAs from Ref.~\newcite{Chi2012} on the right. In the upper right panel, colors indicate a temperature evolution from \unit[2]{K} (black) to \unit[16.5]{K} (red). The observed  fine structures near $\Delta+ \omega_{\rm res}$ require a significant contribution of inelastic tunneling events. These structures require an electronic pairing mechanism with pairing glue gapped below $T_c$. The signatures of the resonance mode require a sign changing pairing state.}
\label{Figcon2}
\end{figure}
As a specific example, we compare our theory with available experimental data we consider  LiFeAs from Ref.~\newcite{Chi2012,Hess2016} (see right panel of Fig.\ref{Figcon2}, where the black curve is at \unit[2]{K} and should be compared to theoretical predictions in the left panel).
In this case, just like in  many high-temperature superconductors, the electronic spectrum
in the normal state is non-flat. Thus, already in the normal state the  elastic conductance shows a clear energy-dependence
around the Fermi edge. One way to treat this is to normalize the experimental spectra with the
normal state conductance. This normalization was used in Ref.~\newcite{Chi2012}.
In Fig.~\ref{Figcon2}a we plot the normalized conductance $\sigma_{\text{norm}}(U)=\nicefrac{\sigma_{\text{sc}}(U)}{\sigma_{\text{nc}}(\sqrt{U^{2}-(\Delta/e)^{2}})}$,
which facilitates a direct comparison with experiments as the effects
of broken particle-hole symmetry is reduced noticeably.
Even the detailed fine structures above the largest
gap are fully consistent with the behaviour seen in our theory including inelastic tunneling (note
that the experimental peak in the second derivative appears at energy $\approx2.4\Delta$
consistent with a neutron resonance mode below $2\Delta$ \cite{Inosov10,RossatMignod1991,Bourges1997,Yu2009}). 
A significant loss of tunneling spectral weight can be seen in the
superconductor for voltages $eU<\Delta+\omega_{\text{res}}$, especially
at $eU\approx2\Delta$, and a following strong increase of the normalized
conductance due to inelastic contributions from the scattering off
the resonance mode. 
In the normalized second derivative this gives
rise to a peak at $eU\approx\Delta+\omega_{\text{res}}$ as also seen in the experimental data. Note, that for the pure elastic theory one expects a dip at this position. Thus, we conclude that inelastic tunneling is present here and that the pairing state of this system must be sign-changing, corresponding to an unconventional mechanism for superconductivity with a pairing interaction that dramatically reorganizes as one enters the superconducting state.
In addition, the obtained normalized spectrum fits well with tunneling data for various other iron
based superconductors~\newcite{Fasano10,Chi2012,Wang12} and cuprate
superconductors~\newcite{Valles91,Jenkins09,Renner1998,Seidel97},
except for the conductance in the gap region $eU<\Delta$ due to possible
double gaps or nodes of the gap.  

In summary, we demonstrated that a  quantitative description of STM tunneling spectra of unconventional superconductors  requires the inclusion of both, elastic and inelastic tunneling events. In the latter case a tip electron tunnels into an off-shell state and eventually relaxes to the Fermi energy by exciting a collective mode. This has long been demonstrated to be of importance in the normal state. Here we show that inelastic tunneling events are responsible for the frequently observed peak-dip features seen in STM spectra in the superconducting state  of  iron-based and cuprate materials. We utilize the fact that inelastic tunneling is directly related to momentum averaged bosonic excitations and demonstrate specifically that the much discussed system LiFeAs is governed by an electronic pairing mechanism with sign changing gap. To this end we performed explicit calculations of the elastic and inelastic tunneling spectrum for a spin-fluctuation induced pairing state that show striking similarities with the experimental data. Thus,  inelastic tunneling offers a new spectroscopic approach to  identify and constrain the collective modes that are responsible for unconventional superconductivity.

\begin{acknowledgments}
The authors are grateful to R. M. Fernandes, Ch. Hess, and  P. Kumar Nag  for discussions and acknowledge funding by the DFG under the grants SCHM 1031/7-1 and WU 349/12-1. This work was performed in part at the Aspen Center for Physics, which is supported by National Science Foundation grant PHY-1066293.
\end{acknowledgments}


\newpage

\begin{widetext}

\section{Derivation of low-energy tunnel Hamiltonian}

In the following, we develop a low-energy theory describing the tunneling spectra between a superconductor and a normal metal from a microscopic model including \textit{elastic} and \textit{inelastic} tunneling channels. The corresponding low-energy Hamiltonian is of the following form:
\begin{align}
\mathcal H^{\text{eff}}_{\text{tun}} &= \sum_{\vec k, {\vec k'} \atop \sigma,\sigma'} \bigl[ t_{\vec k,\vec k'}^\e  \delta_{\sigma,\sigma'} + \sum_{\vec q} t_{\vec k,\vec k',\vec q}^{\text i,\sigma,\sigma',n} \cdot \hat \Phi_{\vec q}^n \bigr] \hat c_{\vec k, \sigma,s}^\dagger \hat c_{\vec k', \sigma',t}  \, ,   \label{1}
\end{align}

\noindent where $t_{\vec k,\vec k'}^\e$ and $t_{\vec k,\vec k',\vec q}^{\text i,\sigma,\sigma',n}$ are the elastic and inelastic tunneling matrix elements for electrons tunneling between the superconductor and the normal tip of an scanning tunneling microscope. $\vec{k}$ and $\sigma$ are the wave vector and spin degrees of freedom. For simplicity, we omit band indices. The operators $\hat \Phi_{\vec q}^n$ with mode index $n=1,\ldots, \mathcal N$ depicts the relevant collective bosonic degrees of freedom in the superconductor, whose dynamics are governed by the bare boson Hamiltonian $\mathcal H_{\Phi,0}$. Finally we define the creation operator  $\hat c^\dagger$ for the electrons in the superconductor $s$ and the tip $t$.

We obtain this low energy Hamiltonian from the generic, \textit{purely elastic} high-energy tunnel Hamiltonian $\mathcal H =\mathcal H_0 +\mathcal H_{\text{tun}}^0$ including a Yukawa-like electron-boson coupling
\begin{align}
\mathcal H_0 &= \sum_{\vec k,\sigma=\uparrow,\downarrow }  \bigl[ \epsilon_{\vec k}^s \hat c_{\vec k,\sigma,s}^\dagger \hat c_{\vec k,\sigma,s} +  \epsilon_{\vec k}^t \hat c_{\vec k,\sigma,t}^\dagger \hat c_{\vec k,\sigma,t}\bigr]  +\sum_{\vec k, \vec q \atop \sigma,\sigma', n} \hat  c_{\vec k+\vec q, \sigma,s}^\dagger \alpha_{\vec q,\sigma,\sigma'}^n \hat c_{\vec k,\sigma',s} \hat\Phi_{\vec q}^n + \mathcal H_{\Phi,0} + \mathcal H_{\text{el-el}} \, ,\label{2} \\
\mathcal H_{\text{tun}}^0&= \sum_{\vec k,\vec k' \atop \sigma} \bigl[t_{\vec k,\vec k'}^\e \hat c_{\vec k, \sigma,s}^\dagger \hat c_{\vec k', \sigma,t} + \text{h.c.}  \bigr]   \, . \label{3}
\end{align}

In the following, we will integrate out the high-energy degrees of freedom in order ot derive an effective low-energy theory. Let us therefore  first write down the corresponding action of the above Hamiltonian
\begin{align}
S =S_{\Phi,0} + \int_{k,p}\hat \psi_k^\dagger \bigl[ \hat G_{k,0}^{-1} \delta_{k,k'} +  \hat \Lambda_{k,k'}   \bigr]  \hat \psi_{k'}
\end{align}
with the bare propagator and interaction matrix
\begin{align}
\begin{split}
\hat G_{k,0}^{-1} &= \text{diag}\bigl([G_{k,0}^s]^{-1}, [G_{k,0}^t]^{-1}   \bigr) \,   \\
\hat \Lambda_{k,k'} &= - \mx{ \sum_n \alpha_{k,k'}^n \Phi_{k-k'}^n & t_{k,k'}^\e \\ \bar t_{k,k'}^\e & 0   }   \, .
\end{split}
\end{align}

Here, we do not consider the repulsive electron-electron interaction, which will be renormalized in the usual way~\cite{Morel62} for the low-energy theory. To derive this effective theory one separates the fermionic fields into low-energy modes and high-energy modes with respect to the smaller momentum cutoff $k_<$ of the low-energy theory
\begin{align}
\hat  \psi_k &= \mehrzeile{ \hat \psi_k^< & \hspace{1cm} \text{, for } \abs{\vec k- \vec k_F}<k_<   \\ 
 \hat \psi_k^> & \hspace{1cm} \text{, for } \abs{\vec k- \vec k_F}>k_<    }  \, . \label{rg3}
\end{align}
Note, that here we assume the momentum cutoff for the tip and superconductor to be the same. We could easily assume different momentum cutoffs, but as we will later see the high-energy states of the tip will have no influence on the low-energy theory. The action can  conveniently be rewritten as
\begin{align}
S =S_{\Phi,0} + \int_{k,k'}   \mx{\hat \psi_k^< \\ \hat \psi_k^>}^\dagger  \mx{ [\hat G_{k,k'}^{<<}]^{-1}    & \hat \Lambda_{k,k'}  \\ \hat \Lambda_{k,k'} & [\hat G_{k,k'}^{>>}]^{-1}   }   \mx{\hat \psi_{k'}^< \\ \hat \psi_{k'}^>}  \, .
\end{align}
The bare dynamics of the low-energy and high-energy sectors are given by the $\hat G_{k,k'}^{<<}, \hat G_{k,k'}^{>>}$ propagators
\begin{align}
\begin{split}
[\hat G_{k,k'}^{<<}]^{-1} &= [\hat G_{k,0}^< ]^{-1} \delta_{k,k'} + \hat \Lambda_{k,k'}  \, ,\\ 
[\hat G_{k,k'}^{>>}]^{-1} &= [\hat G_{k,0}^> ]^{-1} \delta_{k,k'} + \hat \Lambda_{k,k'}  \, ,
\end{split}   \label{rg5}
\end{align}
with the free high- and low-energy propagators
\begin{align*}
[\hat G_{k,0}^> ]^{-1}&=\hat G_{k,0}^{-1}  \cdot \theta(\abs{\vec k- \vec k_F}-k_<)\, , \\
[\hat G_{k,0}^< ]^{-1}&=\hat G_{k,0}^{-1}  \cdot \theta(k_<-\abs{\vec k- \vec k_F})\, .
\end{align*}
In contrast, the non-diagonal elements $\hat \Lambda_{k,k'}$ couple the $\hat \psi^>$ and $\hat \psi^<$ fields via the scattering with bosons and tunneling processes between the tip and the superconductor. Using the usual identities of Gaussian integrals we can integrate out the $\hat \psi^>$ field
\begin{align}
 \int D[(\hat \psi^>)^\dagger, \hat \psi^>] e^{-S} &= e^{-S_{\text{eff}}}   \label{rg6}
\end{align}
with the effective action
\begin{align}
S_{\text{eff}}&=S_{\Phi,0} - \tr \ln [\hat G^{>>}]    \label{rg7}  \\
&+ \int_{k,k'}   \hat \psi_k^< \biggl( [\hat G_{k,k'}^{<<}]^{-1} - \int_{p,p'}    \hat \Lambda_{k,p} \hat G_{p,p'}^{>>}  \Lambda_{p',k'}  \biggr)\hat \psi_{k'}^<   \, . \nonumber
\end{align}

\begin{figure}
\centering
\includegraphics[width=0.49\textwidth]{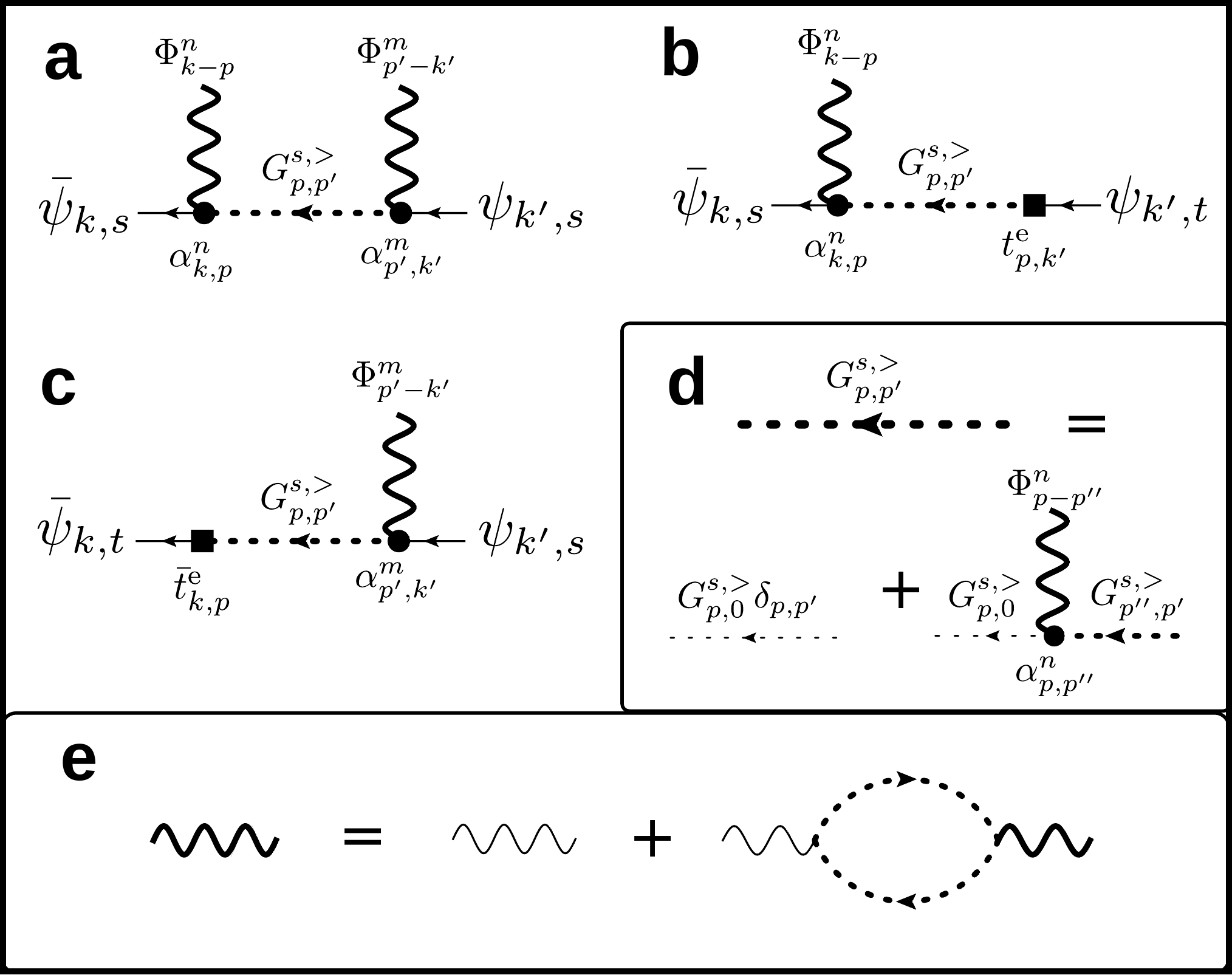}
\caption{\textbf{Interaction vertices generated by integrating out the high-energy fermions.} The diagram \textbf{a} gives describes scattering processes including an virtual off-shell state. The diagrams \textbf{b} and \textbf{c} describe the additional inelastic tunneling matrix element that emerge in the low-energy theory involving the high-energy propagator shown in \textbf{d}. The graph \textbf{e} describes the renormalization of the phonons in the superconductor due to the high-energy fermionic quasiparticles.}
\label{effaction}
\end{figure}

The trace-log term will give rise to a bosonic self-energy (see polarization bubble in Fig.~\ref{effaction}(e))
\begin{align}
 \tr \ln [\hat G^{>>}] &= \frac{1}{2} \int_{\vec q} \Phi_{k-k'}^n \Phi_{-k+k'}^m \int_k \tr \bigl[ \hat G_{k,0}^> \alpha_{k,k'}^n \hat G_{k',0}^>  \alpha_{k',k}^m   \bigr] + \mathcal{O}(\phi^4) \,,   \label{rg8}
\end{align}
which just describes the renormalization of the low-energy phonons due to the presence of the high-energy fermions. Importantly, the polarization operator is not depending on the elastic tunneling elements $t^\e$, which only occur in the $\phi^4$ terms and therefore the screening is in leading order not affected by the presence of the tunneling term in the Hamiltonian.\\
The other additional contribution for the low-energy theory is the last term in~\zref{rg7}. For the evaluation we need to inverse of the second equation in~\zref{rg5}, which has the formal solution
\begin{align}
\hat G_{k,k'}^{>>} &= \hat G_{k,0}^>  \delta_{k,k'} + \int_p \hat G_{k,0}^> \hat \Lambda_{k,p} \hat G_{p,k'}^{>>} = \mx{\hat G_{k,k'}^{>>,ss}  & \hat G_{k,k'}^{>>,st}  \\ \hat G_{k,k'}^{>>,ts}  & \hat G_{k,k'}^{>>,tt} }  \, .  \label{rg9}
\end{align}
As can be easily seen from this expression, the propagators $ \hat G_{k,k'}^{>>,st},  \hat G_{k,k'}^{>>,ts} \sim t^\e$ vanish for the uncoupled system. Using this property of $\hat G_{k,k'}^{>>}$ it is straightforward to expand
\begin{align}
  &   \hat \Lambda_{k,p} \hat G_{p,p'}^{>>}  \Lambda_{p',k'} = G_{p,p'}^{>>,ss}\bigr|_{t^\e=0}    \,  \mx{\sum_{n,m} \alpha_{k,p}^n \Phi_{k-p}^n \alpha_{p',k'}^m \Phi_{p'-k'}^m & \sum_n \alpha_{k,p}^n \Phi_{k-p}^n  t_{p,k'}^\e \\  \sum_n t_{k,p}^\e \alpha_{p',k'}^n \Phi_{p'-k'}^n & 0 }  + \mathcal{O}[(t^\e)^2]    \label{rg10}
\end{align}
in leading order in $t^\e$. The diagonal term describes the scattering from low-energy fermions to off-shell states and vice versa via a boson exchange, which could be described via an new effective electron-boson vertex, see Fig.~\ref{effaction}(a). In the following, we will neglect this term as it should only give us important corrections for our fermionic quasiparticles for high-energies and not for the low-energy theory since here the dominant processes are the scattering processes between on-shell electrons. In contrast, the off-diagonal terms shown in Fig.~\ref{effaction}(b) and (c) give us new relevant physics: The tunneling process from a tip to an off-shell state in the superconductor and from there an inelastic scattering to a state near the Fermi surface via the excitation/absorption of a boson and vice versa. \\
From \zref{rg9} it is clear that the Dyson equation for $ G_{k,k'}^{>>,ss}\bigr|_{t^\e=0}$ reads
\begin{align}
 G_{k,k'}^{>>,ss}\bigr|_{t^\e=0} &=  \hat G_{k,0}^{s,>} \delta_{k,k'} +  \int_p  G_{k,0}^{s,>} \sum_n \alpha_{k,p}^n \Phi_{k-p}^n  G_{p,k'}^{>>,ss}      \nonumber \\
 & \approx \frac{\theta(\abs{\vec k}-k_<)}{-D}  \biggl[ \delta_{k,k'} + \int_p \sum_n   \alpha_{k,p}^n \Phi_{k-p}^n  G_{p,k'}^{>>,ss}      \biggr]      \nonumber \\
 &= \frac{\theta(\abs{\vec k}-k_<)}{-D}  \biggl[ \delta_{k,k'} + \sum_n \frac{  \alpha_{k,k'}^n \Phi_{k-k'}^n}{-D} + \int_{p^>}\sum_{n,m} \frac{  \alpha_{k,p}^n \Phi_{k-p}^n  \alpha_{p,k'}^m \Phi_{p-k'}^m}{(-D)^2}  +\ldots  \biggr] \label{rg11}
 \end{align} 
In the end we  approximated the bare off-shell propagators $ \hat G_{k,0}^{s,>} \approx - 1/D\,  \theta(\abs{\vec k}-k_<)$, where $D$ is the band-width of the system, as explained in the main text. Inserting \zref{rg11} into \zref{rg10} we find the following effective low-energy action~\zref{rg7}
\begin{align}
S_{\text{eff}}&= \overbrace{S_{\Phi}^{\text{screen}}  + S_{\text{el}}^< + S_{el-\Phi}}^{S_0^{\text{screen}}}\nonumber   \\
& - \biggl(\int \limits_{k^<, k'^<} \bigl[t_{k,k'}^\e + \int \limits_{p^>} \sum_n \alpha_{k,p}^n \Phi_{k-p}^n  t_{p,k'}^\e +\ldots\bigr] \bar \psi_{k,s} \psi_{k',t}  + \text{h.c.}  \biggr)   \label{rg12}
\end{align}
The $S_0^{\text{screen}}$ described the low energy electrons (and bosons) in the tip and the superconductor in the absence of tunneling. This system can be calculated using the usual field-theoretical models like the Midgal-Eliashberg theory. The low-energy tunneling part of the action has not only an elastic part, but also acquires  additional one- and multiple-boson inelastic channels. Rewriting this in terms of an Hamiltonian, we end up with the low-energy tunneling Hamiltonian given in the main text.

\section{Derivation of the tunneling current for the spin-fermion model}
Performing usual perturbation theory to leading order in $t^{\e}$, the elastic contribution\cite{Jandke15} of the STM current into a superconductor is just given by the usual Landauer-B\"uttinger expression (note that we defined the bias here $U=\mu_T-\mu_T$)
\begin{align}
 I^\text e (U)&=  4 \pi e \abs{t^\e}^2 \intinfty d\omega \bigl[ n_F(\omega-eU)-n_F(\omega) \bigr]\nu_S(\omega) \nu_T(\omega-eU)
 \end{align} 
where we used the approximation $t_{\vec k,\vec k'}^\e \approx t^\e$, and $\nu_{S/T}$ are the fermionic DOS in the superconductor/tip. 
The inelastic current can be directly computed from the many-body Hamiltonian in the main text using  field-theoretical methods~\cite{Jandke16} and is given by
\begin{align}
I^\text{i}(U) &=\frac{4 \pi e \abs{t^\text{i}}^2}{\nu_S^0} \intinfty d\omega_1 d\omega_2 \nonumber \\
& \hspace{1cm}\biggl(  \alpha^2 F_{\text{tun}}(\omega_1) \,  \nu_S(\omega_2) \nu_T(\omega_2-\omega_1-eU)  \nonumber \\
&\hspace{1cm}\biggl[ n_F(\omega_2-\omega_1-eU) n_B(\omega_1) \bigl[1-n_F(\omega_2) \bigr]    - n_F(\omega_2) \bigl[1+n_B(\omega_1) \bigr] [1-n_F(\omega_2-\omega_1-eU) \biggr] \nonumber \\
& \hspace{1cm} +   \alpha^2 F_{\text{tun}}(\omega_1) \,  \nu_S(\omega_2) \nu_T(\omega_2+\omega_1-eU)  \nonumber \\
 & \hspace{1cm}    \biggl[ n_F(\omega_2+\omega_1-eU) \bigl[1+n_B(\omega_1)\bigr] \bigl[1-n_F(\omega_2) \bigr] - n_F(\omega_2) n_B(\omega_1) [1-n_F(\omega_2+\omega_1-eU)] \biggr]  \biggr)  \nonumber
\end{align}
where we defined $\nu_S^0$ as the normal state DOS of the superconductor at the Fermi surface and
\begin{align}
 \alpha^2 F_{\text{tun}}(\omega) &= \frac{\nu_S^0}{2}\sum_{\vec q, n,m \atop \sigma,\sigma'} \alpha_{\vec q, \sigma,\sigma'}^n \bigl[\alpha_{\vec q,\sigma',\sigma}^m\bigr]^* \frac{\chi_{\vec q,nm}^R(\omega)- \chi_{\vec q,nm}^A(\omega) }{-2\pi i}	
\end{align}
and $\chi_{\vec q,nm}^R(t,t') =- i \theta(t-t')  \erw{\bigl[\hat \Phi_{\vec q}^n(t) ,\hat  \Phi_{-\vec q}^m(t')\bigr]}$ is the retarded boson propagator of the superconducting system (of the effective low-energy theory). \\
For the spin fermion-model it holds $\alpha_{\vec q, \sigma,\sigma'}^n = g_{\vec q} \tau_{\sigma,\sigma'}^n$ and $\chi_{\vec q,nm}^R(t,t') = \chi_{\vec q}^R(t,t') \delta_{nm}$ such that we find here the bosonic tunneling spectrum $\alpha^2 F_{\text{tun}}(\omega) = g^2 \chi(\omega)$ as the integrated spin susceptibility
\begin{align}
g^2 \chi(\omega) &= \frac{\nu_S^0}{2}\sum_{\vec q, n,m \atop \sigma,\sigma'}  \abs{g_{\vec q}^2} \tau_{\sigma,\sigma'}^n \tau_{\sigma',\sigma}^m\, \delta_{n,m} \frac{\chi_{\vec q}^R(\omega)- \chi_{\vec q}^A(\omega) }{-2\pi i}	   \nonumber \\
&= 3 \nu_S^0 \sum_{\vec q}  \abs{g_{\vec q}^2} \frac{\Im \chi_{\vec q}^A(\omega)}{\pi}
\end{align}
In the limit of a constant tip DOS, we can then easily derive the expressions for the differential conductance given in the main text.

\end{widetext}

\end{document}